\documentclass[aps, showpacs,prb,twocolumn,reprint]{revtex4-1}

\usepackage{amsmath}           
\usepackage{amsfonts}         
\usepackage{amssymb}          
\usepackage{graphicx} 
\usepackage{stackrel}
\usepackage{paralist}
\usepackage{xcolor}
\usepackage{ulem}
\usepackage{subfigure}

\newcommand{\I}{\mathrm{i}}
\newcommand{\refEq}[1]{Eq. (\ref{#1})}
\newcommand{\refFig}[1]{Fig. \ref{#1}}

\DeclareMathOperator{\Li}{Li}

\def\Xint#1{\mathchoice
{\XXint\displaystyle\textstyle{#1}} 
{\XXint\textstyle\scriptstyle{#1}} 
{\XXint\scriptstyle\scriptscriptstyle{#1}} 
{\XXint\scriptscriptstyle\scriptscriptstyle{#1}} 
\!\int}
\def\XXint#1#2#3{{\setbox0=\hbox{$#1{#2#3}{\int}$ }
\vcenter{\hbox{$#2#3$ }}\kern-.5\wd0}}
\def\Pint{\Xint{P}}

\begin{document}
\title{Direct Proportionality between the Kondo Cloud and Current Cross Correlations in Helical Liquids}

\author{Thore Posske}
\author{Bj{\"o}rn Trauzettel}
\affiliation{Institute for Theoretical Physics and Astrophysics, University of W{\"u}rzburg, D-97074 W{\"u}rzburg, Germany}



\begin{abstract}
The Kondo cloud, represented by the correlations between the magnetic moment and the spin density in the leads of a Kondo setup,
is until now eluding its observation. 
We exploit the unique coupling of spin and direction of motion of the excitations in the recently discovered helical liquids in a setup with two leads to establish a proportionality between the Kondo cloud and the time resolved current cross correlations.
This relation holds around a specific choice of model parameters.
Thereby, we propose a direct way to detect the Kondo cloud in a least invasive manner since the current cross correlations are measurable far away from the magnetic moment.
We, furthermore, discuss how our predictions are modified 
if the model parameters are varied away from the specific choice.
\end{abstract}

\maketitle

\section{Introduction}
A magnetic moment coupled to a fermionic bath, a so-called Kondo setup, is one of the most basic combinations of two quantum mechanical systems with spin. Yet its description has attracted the attention of theoretical as well as experimental physicists already for many decades \cite{DeHaasDeBoer1934, KondoSeminalPaper,HewsonTheKondoProblemToHeavyFermions}. 
A general feature of a Kondo setup is that at temperatures smaller than the Kondo temperature, the excitations of the fermionic bath orientate their spin to screen the magnetic moment. This screening ''cloud" of excitations has been termed the Kondo cloud\cite{MuellerHartmann1969SpinCorrelationInDiluteMagneticAlloys,BordaKondoScreeningCloudInAOneDimensionalWire,SimoninLookingForTheKondoCloud, Affleck2009TheKondoScreeningCloudWhatIsIt}, and its extent is considerably long-ranged.
A concrete observable quantifying the presence of the Kondo cloud at a given position in the fermionic bath is the equal time correlator between the spatially dependent spin density in the bath and the spin of the magnetic moment. In fact, we are going to use this correlator as a synonym for the Kondo cloud throughout this article.
All attempts to measure the Kondo cloud have failed so far.
For instance, there appear diverse technical and conceptual problems if it is attempted to directly measure the Kondo cloud with two spin-sensitive scanning tunneling microscopy (STM) tips \cite{BordaKondoScreeningCloudInAOneDimensionalWire}. Maybe the most conceptually problematic aspect of that strategy is that the coupling of a spin sensitive STM tip to the magnetic moment is likely to influence the Kondo effect and hence the Kondo cloud itself.
In this regard, a more promising experiment to measure the Kondo cloud should employ an indirect measurement that does not affect the 
close vicinity of the magnetic moment \cite{Affleck2009TheKondoScreeningCloudWhatIsIt}.
Along these lines, recent suggestions propose to first find an observable that alters its behavior decisively upon reaching the Kondo length  \cite{AffleckSimonDetectingTheKondoScreeningCloudAroundAQuantumDot,PereiraRodrigoLaflorencieAffleckHalperin2008KondoScreeningCloudAndChargeStaircaseInOneDimensionalMesoscopicDevices,OregHowToMeasureDirectlyAKondoCloudsLength}. 
Notably, the authors of Ref. \onlinecite{PattonHafermannBrenerLichtensteinKatsnelson2009ProbingTheKondoScreeningCloudViaTunnelingCurrentConductanceFluctuations} mention a visible signature of the Kondo cloud in conductance fluctuations close to the magnetic moment.

New light on the Kondo cloud can be shed  by means of the recently discovered quantum spin Hall insulator \cite{KoenigQSHE,BernevigZhangQSHEInHgTeWells,KaneMeleQSHEGraphene}. This quantum phase accommodates a metallic edge state, the helical liquid, that is protected from inelastic backscattering by time reversal symmetry. 
The unique feature of the helical liquid is the direct coupling between the direction of motion and the spin of its excitations. Excitations with spin down traverse the edges clockwise, while excitations with spin up propagate conversely.
In this way, the information about interaction processes
with a magnetic moment is carried away from the scattering region in a spin resolved fashion.

Systems, where a magnetic moment is coupled to helical liquid leads, have already been analyzed
\cite{MaciejkoLiuOregQiQuZhangKondoEffectInTheHelicalEdgeLiquidOfTheQuantumSpinHallState, LawSengQuantumDotIn2DimTopologicalInsulator, Maciejko2012KondoLatticeOnTheEdgeOfATwoDimensionalTopologicalInsulator}. 
However, the focus never lay on the Kondo cloud.
In Ref. \onlinecite{PosskeLiuBudichTrauzettel2012ExactResultsForTheKondoCouldOfTwoHelicalLiquids},
we have discovered representative parameter configurations, the Toulouse points \cite{Toulouse1969ExactExpressionOfEnergyOfKondoHamiltonianBaseStateForAParticularJZValue,EmeryKivelsonKondo}, for which the corresponding Hamiltonian is mappable to a quadratic one. This mapping enabled us to calculate the shape and non-equilibrium properties of the Kondo cloud non-perturbatively.

In this article, we improve substantially upon existing suggestions of measuring the Kondo cloud by 
establishing a close relation of the Kondo cloud to  
time- and space-resolved current cross correlations in the leads measured sufficiently far away from the magnetic moment. The measuring distance should be larger than the Kondo length and is, in theory, not limited from above. In praxis, the purity of the sample, temperature, etc. give an upper bound.
Our suggestion is based on a setup consisting of a magnetic moment of spin $\hbar/2$ coupled to two strongly interacting helical liquid leads (see \refFig{figSetup}), with the Luttinger parameters $g_\text{t}$ and $g_\text{b}$ both being $1/2$ (see Section \ref{sctnModel}). 
In this case, a two-channel Kondo Hamiltonian is appropriate to describe our system \cite{LawSengQuantumDotIn2DimTopologicalInsulator}.
By propagating the density operator backwards in time, we relate the current cross correlations to the Kondo cloud in an exact manner for a broad range of the Kondo parameters. In first order of the $z$-couplings, this relation is a direct proportionality.
As a concrete example, we give the explicit analytical form of both quantities for vanishing $z$-couplings and emphasize the necessity to measure the current cross correlations in the time domain.
The particular system of helical liquid leads with $g_\text{t} = g_\text{b} = \frac{1}{2}$ is mainly considered here because of the direct and unperturbed appearance of the Kondo cloud in the current cross correlations.
We discuss, finally, the fact that by departing from this particular value on the line $g_\text{t}+g_\text{b} = 1$, 
the signature of the Kondo cloud remains in principle visible in the current cross correlations.

The article is organized as follows. In Section \ref{sctnModel}, we introduce our model and transform the Hamiltonian to an interacting resonant level model. Afterwards, in Section \ref{sctnResults} we derive the proportionality relation between the Kondo cloud and the current cross correlations and present the Fourier transforms of both quantities. In Section \ref{sctnDiscussion}, we discuss how the proportionality is affected by a change of the Luttinger parameters in the leads. Finally, we conclude in Section \ref{sctnSummary} and address some technical details in the Appendix.

\section{Model \label{sctnModel}}
\begin{figure}
\includegraphics[width= .75 \linewidth]{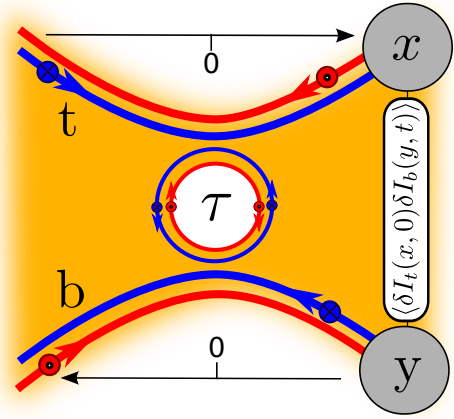}	
\caption{
\label{figSetup}
(Color online) Two helical liquids are coupled to a magnetic moment $\tau$ of spin $\hbar/ 2$.
The blue (red) lines indicate edge channels with spin down (up) moving clockwise (counterclockwise) at the outer edge of the system.
Contacts at $x$ and $y$ allow measurement of the space- and time-resolved current cross correlations.
}
\end{figure}

We consider a magnetic moment of spin $\hbar/2 $ that is coupled to two helical liquids. In our setup shown in \refFig{figSetup}, we assume that the magnetic moment can be realized
by an anti-dot \cite{PosskeLiuBudichTrauzettel2012ExactResultsForTheKondoCouldOfTwoHelicalLiquids,DelcettoCavaliereFerraroSassetti2013GeneratingAndControllingSpinPolarizedCurrentsInducedByAQSHI}.
The black arrows indicate that space is measured along the path of the spin-down excitation separately at each side.
Contacts at the top (t) and the bottom (b) lead allow us to measure the space- and time-resolved current cross correlations $\langle \delta I_\text{t}(x) \delta I_\text{b}(y,t) \rangle$ between the positions $x$ and $y$ with the time delay $t$. 
Here, $\delta A = A - \langle A \rangle$ and $\langle \dots \rangle$ denotes the thermodynamic expectation value. 
For reasons which will becomes clear in Section \ref{sctnResults}, $x$ and $y$ should be of the order of or larger than the Kondo length scale $\lambda^K$ and $t$ of the order of or larger than the Kondo time scale $\tau^K$. Both scales are canonically derived from the Kondo temperature $T^K$ by $\lambda^K = v \tau^K = \frac{v \hbar}{k_B T^K}$; $v$ is the Luttinger parameter of the leads that describes the velocity of their excitations.
Additionally, to resolve the structure of the Kondo cloud spatially, it is essential that either the time delay $t$ is freely tunable or one of the contacts is movable.

The setup is modeled by a two channel Kondo Hamiltonian, as discussed in Refs. \onlinecite{SchillerHershfieldKondo,LawSengQuantumDotIn2DimTopologicalInsulator, PosskeLiuBudichTrauzettel2012ExactResultsForTheKondoCouldOfTwoHelicalLiquids},
with
\begin{align}
\tilde{\mathcal{H}}_a =& \sum_{\sigma \in \{\uparrow, \downarrow\}} \int dx \left(
v_{F,a} \tilde{\Psi}_{a,\sigma}^\dagger(x)(\sigma \I \partial_x)\tilde{\Psi}_{a,\sigma}  	(x)
\right.
\nonumber
\\
&\left. \ \ \ \ \ \ \ + \frac{g_{4,a}}{2} \tilde{\rho}_{a,\sigma}^2(x) + g_{2,a} \tilde{\rho}_{a,\downarrow}(x) \tilde{\rho}_{a, \uparrow}(x)
\right),
\nonumber
\\
\tilde{\mathcal{H}}^K_a =& \sum_{\lambda \in \{x,y,z\}} J^\lambda \tilde{S}_a^\lambda(0) \tau^\lambda.
\label{eqBasicHamiltonian}
\end{align}  
for each lead $a \in \{\text{t},\text{b}\}$.
Here, $\tilde{\mathcal{H}}_a$ describes the helical liquid of lead $a$ with the fermionic fields $\tilde{\Psi}_{a,\sigma}$ of spin $\sigma$.
The real constants are  the Fermi velocity $v_{F,a}$ and $g_{2/4,a}$, the interaction strengths in the leads.
It is convenient to introduce the Luttinger parameters $g_a = \sqrt{v_{J,a}/v_{N,a}}$
and
$v_a = \sqrt{v_{J,a} v_{N,a}}$ instead, where $v_{N/J,a} = v_{F,a} + \frac{g_{4,a}\pm g_{2,a}}{2\pi \hbar}$  \cite{SchoenhammerBosonization}.
The interaction of lead $a$ with the magnetic moment $\tau$ is given by $\mathcal{H}^K_a$. Further, $\tilde{S}^\lambda_a= \frac{\hbar}{2}\sum_{\sigma,\sigma^\prime}\tilde{\Psi}_{a,\sigma}^\dagger \sigma^\lambda_{\sigma,\sigma^\prime}\tilde{\Psi}_{a,\sigma^\prime}$ denotes the spatially resolved spin density in lead $a$ with $\sigma^\lambda$ being the $\lambda$-th Pauli matrix.
We consider isotropic interactions in the $x$ and $y$ directions, i.e., $J^x_a = J^y_a =: J^\perp_a$, and focus on the special case of $g_\text{t}=g_\text{b}=\frac{1}{2}$ and $v_\text{t} = v_\text{b} := v$. In this case, the Hamiltonian is simplified by the application of bosonization and refermionization. 
In fact, the entire line $g_\text{t}+g_\text{b}=1$ can be treated by performing an intermediate Emery-Kivelson rotation, which we address in Section \ref{sctnDiscussion}. 
Details of the procedure are discussed in Refs. \onlinecite{Toulouse1969ExactExpressionOfEnergyOfKondoHamiltonianBaseStateForAParticularJZValue,EmeryKivelsonKondo,SchillerHershfieldKondo, PosskeLiuBudichTrauzettel2012ExactResultsForTheKondoCouldOfTwoHelicalLiquids}.
The refermionized Hamiltonian then becomes an interacting resonant level model
\begin{align}
\mathcal{H}^0 &= 
\sum_{j \in \{\text{t},\hat{\text{t}},\text{b},\hat{\text{b}}\}} 
\int dx \ 
\hbar v \ \Psi^\dagger(x)_{j}(- \I \partial_x) \Psi(x)_{j} 	
,
\nonumber
\\
\mathcal{H}^\perp_{K} &= j^\perp_\text{t} \Psi_\text{t}^\dagger(0)\tau^- + j^\perp_\text{b} \Psi_\text{b}^\dagger(0)\tau^- + H.c.,
\nonumber
\\
\mathcal{H}^z_{K} &= j_\text{t}^z \Psi^\dagger_\text{t}(0) \Psi_\text{t}(0) \tau^z + j_\text{b}^z \Psi^\dagger_\text{b}(0)\Psi_\text{b}(0) \tau^z.
\label{eqSimplifiedHamiltonian}
\end{align}
Notable simplifications compared to the physical Hamiltonian are that two of the former four fermionic fields decouple and the only two-particle term appears in $\mathcal{H}^z_K$.
In \refEq{eqSimplifiedHamiltonian}, the fermionic fields $\Psi$ are non-linear combinations of the physical ones $\tilde{\Psi}$, but the relation that is important for this work is the linear dependence of the transformed densities 
$
\rho_{a}(x) 
  = \frac{1}{4}
    \left(
	3,1,-1,-3
    \right) \times 
$ 
$
    \left(
	\tilde{\rho}_{a,\uparrow}(x),\tilde{\rho}_{a,\downarrow}(x),\tilde{\rho}_{a,\uparrow}(-x),\tilde{\rho}_{a,\downarrow}(-x)
    \right)^T
$
on the physical ones.
The coupling constants are
$j^\perp_\text{t/b} = \frac{J_{\text{t/b}}^\perp \hbar v}{2\sqrt{2 \pi a_c}}$,
where $a_c$ is the cutoff length scale of the bosonization procedure, and $j_\text{t/b}^z = \hbar J^z_\text{t/b}$.
For $j^z_\text{t} = j^z_\text{b}= 0$, the Hamiltonian reaches a Toulouse point \cite{Toulouse1969ExactExpressionOfEnergyOfKondoHamiltonianBaseStateForAParticularJZValue, EmeryKivelsonKondo, SchillerHershfieldKondo, PosskeLiuBudichTrauzettel2012ExactResultsForTheKondoCouldOfTwoHelicalLiquids} where many observables can be calculated analytically.
It is interesting in this context is that the interacting resonant level model has recently attracted new attention because of the development of exact methods at finite temperature and out of equilibrium \cite{Doyon2007NewMethodForStudyingSteadyStatesInQuantumImpurityProblemsTheInteractingResonantLevelModel,MehtaAndrei2006NonequilibriumTransportInQuantumImpurityModelsTheBetheAnsatyForOpenSystems,BoulatSaleur2008ExactLowTemperatureResultsForTransportPropertiesOfTheInteractingResonantLevelModel,AndergassenPletyukhovSchurichtSchoellerBorda2011RenormalizationGroupAnalysisOfTheIRLMAtFiniteBias} to solve it.
Applied to our setup, these methods could extend the range of exactly solvable parameter configurations considerably.

\section{Results \label{sctnResults}}
The Kondo cloud on side $a$ is defined as the spatially resolved correlation of the $z$ spin density in lead $a$ and the $z$ component of the magnetic moment. 
It takes the form 
\begin{align}
\chi^z_a(x) = \langle  \delta \tilde{S}^z_{a}(x) \delta\tau^z  \rangle.
\label{eqDefinitionKC}
\end{align}
In the following, an additional argument of the operators denotes the time in the Heisenberg picture. If no additional argument is given, we imply that it is equal to zero.
The space- and time-resolved current cross correlations are then given by 
\begin{align}
\chi^{cc}(x,y,t) &= \langle \delta I_\text{t}(x) \delta I_\text{b}(y,t) \rangle
\nonumber
\\
 &= (2 e v)^2 \langle \delta \tilde{S}^z_{\text{t}}(x) \delta \tilde{S}^z_{\text{b}}(y,t) \rangle.
\label{eqDefinitionCCC}
\end{align}
It is here where the very special attribute of the helical liquid, namely that the spin density is proportional to the current, initially connects the two quantities of Eqs. (\ref{eqDefinitionKC}) and (\ref{eqDefinitionCCC}).
Next, we express them in the fields of the simplified Hamiltonian in \refEq{eqSimplifiedHamiltonian}, and exploit its symmetry under simultaneous time-reversal and space inversion.
\footnote{The exploited symmetry of the transformed Hamiltonian in \refEq{eqSimplifiedHamiltonian} originates from the symmetries of the physical Hamiltonian in \refEq{eqBasicHamiltonian}, namely, time-reversal symmetry and the symmetry under simultaneous $z$ spin flip and space inversion.}
For the Kondo cloud, we obtain
\begin{align}
\chi^z_a(x)  = \langle   \delta \rho_a (-|x|)  \delta\tau^z \rangle,
\label{eqSimplifiedKondoCloud}
\end{align}
and the current cross correlations become 
\begin{align}
\chi^{cc}(x,y,t)	= (e v)^2 \sum_{\sigma = \pm} \langle \delta\rho_\text{t}(\sigma |x|,0) \delta\rho_\text{b}(-\sigma |y|,t) \rangle.
\label{eqSimplifiedCurrentCurrentCorrelations}
\end{align}
It is crucial for the derivation of \refEq{eqSimplifiedCurrentCurrentCorrelations} that $\langle \rho_\text{t}(|x|) \rho_\text{b}(|y|,t) \rangle = \langle \rho_\text{t}(-|x|) \rho_\text{b}(-|y|,-t) \rangle^* = 0$ for all times $t$.
The physical reason for the last equality is that excitations in different leads are independent of each other before they can interact at the site of the magnetic moment
\footnote{
The feature that the density correlations at negative spatial values
vanish can be rigorously proven on the basis of the relations between
different correlation functions (retarded, advanced, and Keldysh) in
the Keldysh formalism\cite{rammer2004quantum} also known as RAK rules.
}.

To derive and physically motivate the close relation between the quantities of Eqs. (\ref{eqSimplifiedKondoCloud}) and (\ref{eqSimplifiedCurrentCurrentCorrelations}), we take the density operators with positive spatial argument in the summands of the latter equation and propagate them backwards in time before they have interacted with the magnetic moment. This propagation can be done exactly due to the property $\left( \tau^z \right)^2 = \frac{\hbar^2}{4}$ although the Hamiltonian possesses two-particle terms. Details are given in the Appendix.
The result is
\begin{align}
\rho_a&\left(x,-x/v- \epsilon\right) =
\nonumber
\\
&
  \rho_a(-\epsilon,0)  + \frac{\left( j_a^\perp / v \right)^2}{\hbar  \left( 1+(\frac{j^z_a}{4 v})^2 \right)} (\tau^z(-\epsilon)+ \hbar/2)
  \nonumber
  \\
 &  + \left( \I \frac{j_a^\perp}{\hbar v} \frac{1+\I \frac{j_a^z}{4 v }}{(1-\I \frac{j_a^z}{4 v})^2} \tau^+(-\epsilon) \Psi_a(-v \epsilon,0) + H.c.\right),
\label{eqBackpropagationOfDensity}
\end{align}
for any positive time $\epsilon$. 
It is seen here that the density at positive spatial values carries information about the magnetic moment in the term proportional to $\tau^z$. This is the basic reason that it is possible to measure the Kondo cloud by looking at current cross correlations.
Inserting \refEq{eqBackpropagationOfDensity} into \refEq{eqSimplifiedCurrentCurrentCorrelations}, we encounter three types of expectation values:
\begin{inparaenum}[(i)]
      \item density correlations of the form $\langle \delta\rho_\text{t}(-\epsilon,0) \delta\rho_\text{b}(-|y|,t-|x|-\epsilon)\rangle$, which vanish as explained in the derivation of \refEq{eqSimplifiedCurrentCurrentCorrelations}, 
			\item terms like $\langle \delta\tau^z(-\epsilon) \delta\rho_\text{b}(-|y^\prime|,t-|x|-\epsilon) \rangle$
				that resemble the Kondo cloud, and
      \item correlators similar to
$
	\langle \delta(\tau^+(-\epsilon) \Psi_\text{t}(-\epsilon,0)) \delta\rho_{-a}(-|y|,t-|x|-\epsilon)\rangle,
      \label{eqExampleOfUnwantedCorrelations}
$
			which vanish in first order in $j_\text{t/b}^z$.
\end{inparaenum}
Retaining only the first and zeroth orders in $j^z_\text{t/b}$, we therefore obtain
\begin{align}
\chi^{cc}(x,y,t)	=& \frac{e^2}{\hbar}
		  \left(
		    (j^\perp_\text{t})^2 \langle \delta\rho_\text{b}(-|y|, t-|x|/v) \delta\tau^z \rangle^*
\right.
\nonumber
\\
&\left.
    		    + (j^\perp_\text{b})^2 \langle \delta\rho_\text{t}(-|x|,-t-|y|/v) \delta\tau^z \rangle
		  \right).
\label{eqAnalyticallyProportionality}
\end{align}
This expression consists of two summands each of which already resembles the Kondo cloud defined in \refEq{eqDefinitionKC}.
The aim is now to choose a time frame for which one of the summands becomes proportional to the Kondo cloud and the other one is suppressed.
To reveal the Kondo cloud from the first summand, we constrain $v t - |x| - |y| < 0$ and shift the time argument of the density operator into its spatial argument. The same can be done for the second summand in the case $v t + |x| + |y| > 0$.
For suppressing the summand that fails to be proportional to the Kondo cloud within one of the respective time frames, we argue that there is an intrinsic time scale $\tau^c$ after which 
$\langle \delta \rho_a(-\eta,0) \delta \tau^{z}(t) \rangle$ decays rapidly if $|t|>\tau^c$.
Here, $\eta$ is a finite but small position in space. 
This assumption
is physically motivated by the fact that a scattering problem lacks periodicity and usually exhibits no infinite length correlations. 
We consider the required time 
$\tau^c$ to be of the order of the Kondo time scale $\tau^K$
because it is the largest time scale that is immediately connected to the Hamiltonian.
Nevertheless, the concrete choice of
this time scale
$\tau^c$
is of no significance in principle for the following results.
This argument leads us to two time frames fulfilling the demanded conditions:
(i) $(|x| + |y|)/v > t > \tau^K$, where the current cross correlations are greatly dominated by the first correlator in \refEq{eqAnalyticallyProportionality} since $|- t -(|x| +|y|)/v|$ is more than $2 \tau^K$ larger than 
$ t + (|x| +|y|)/v$.
(ii)  $- \tau^K >  t > -(|x|+|y|)/v$, where the current cross correlations are dominated by the second correlator in \refEq{eqAnalyticallyProportionality}. 
We then find the central result of this article: 
\begin{align}
\chi^{cc}(x,y,t) \approx
\frac{ 2 e^2 v k_B}{\hbar^3} 
\begin{cases}
 T^K_\text{t} \chi^z_\text{b}\left(|x|+|y| - v t \right)	& \text{(i)}
\\
 T^K_\text{b}  \chi^z_\text{t}\left(|x|+|y| + v t \right) & \text{(ii)},
\end{cases}
\label{eqFinalProportionality}
\end{align}
where
$T^K_{\text{t/b}} =\frac{\hbar ( j_\text{t/b}^\perp )^2}{2 k_b v} = \frac{ \hbar (J^\perp_{\text{t}/\text{b}})^2}{16 \pi a_c k_B v}$ is the Kondo temperature of one lead calculated as if the other one does not exist\cite{SchillerHershfieldKondo}.
The deviation of 
\refEq{eqFinalProportionality}
from an identity is suppressed arbitrarily by measuring further away from the magnetic moment, i.e. increasing $|x|+|y|$, and therefore it vanishes for any practical purpose.
Rephrased, \refEq{eqFinalProportionality} states that, for certain time frames, the Kondo clouds of both leads are mirrored in the current cross correlations by a direct proportionality, and the proportionality factors are determined by the Kondo temperature of the respective opposite lead.

\begin{figure}
\subfiguretopcaptrue
\subfigure[\label{figFTKC}]
{
\includegraphics[width = 0.9 \linewidth]{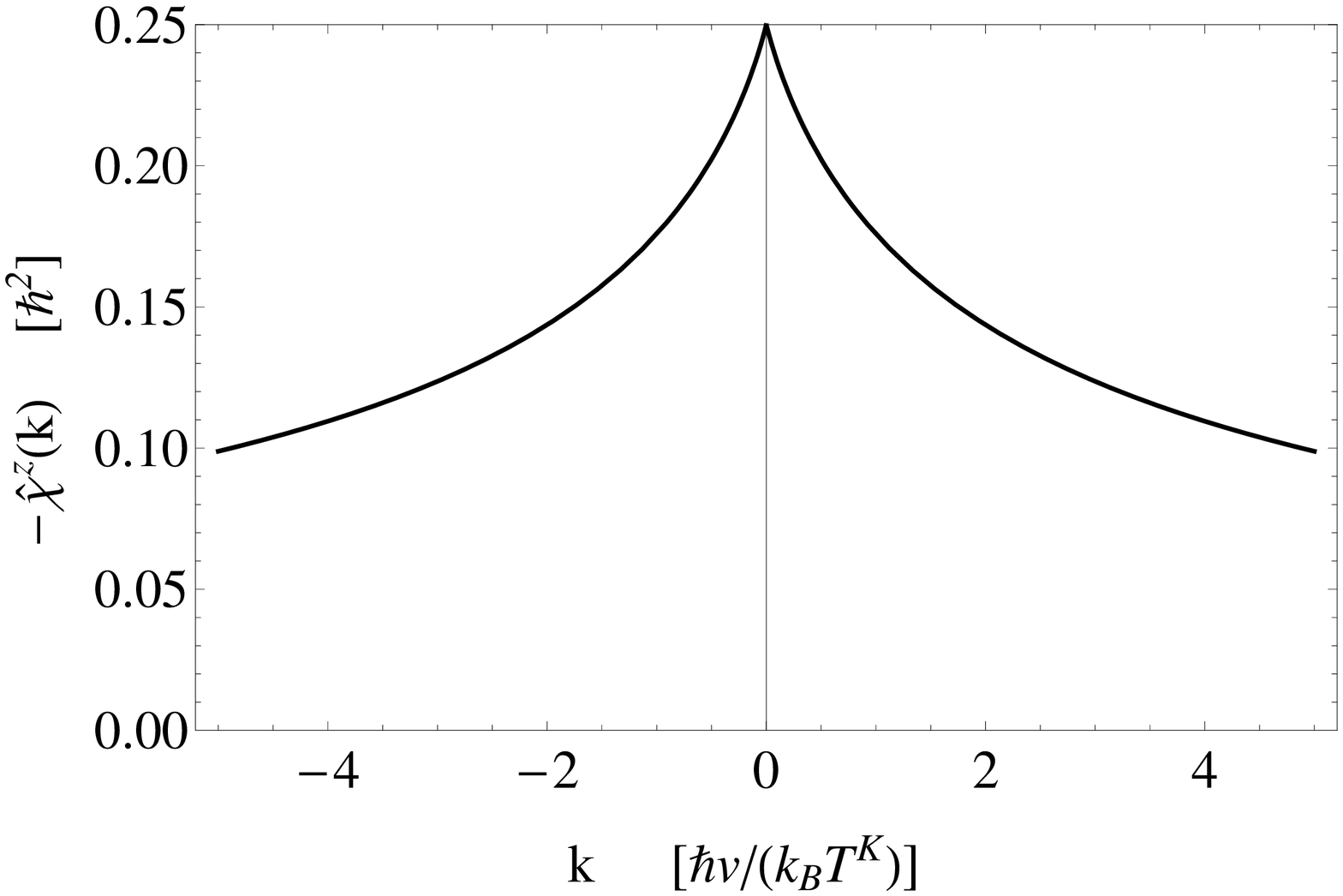}
}

\subfigure[\label{figFTCCC}]
{
\includegraphics[width = 0.9 \linewidth]{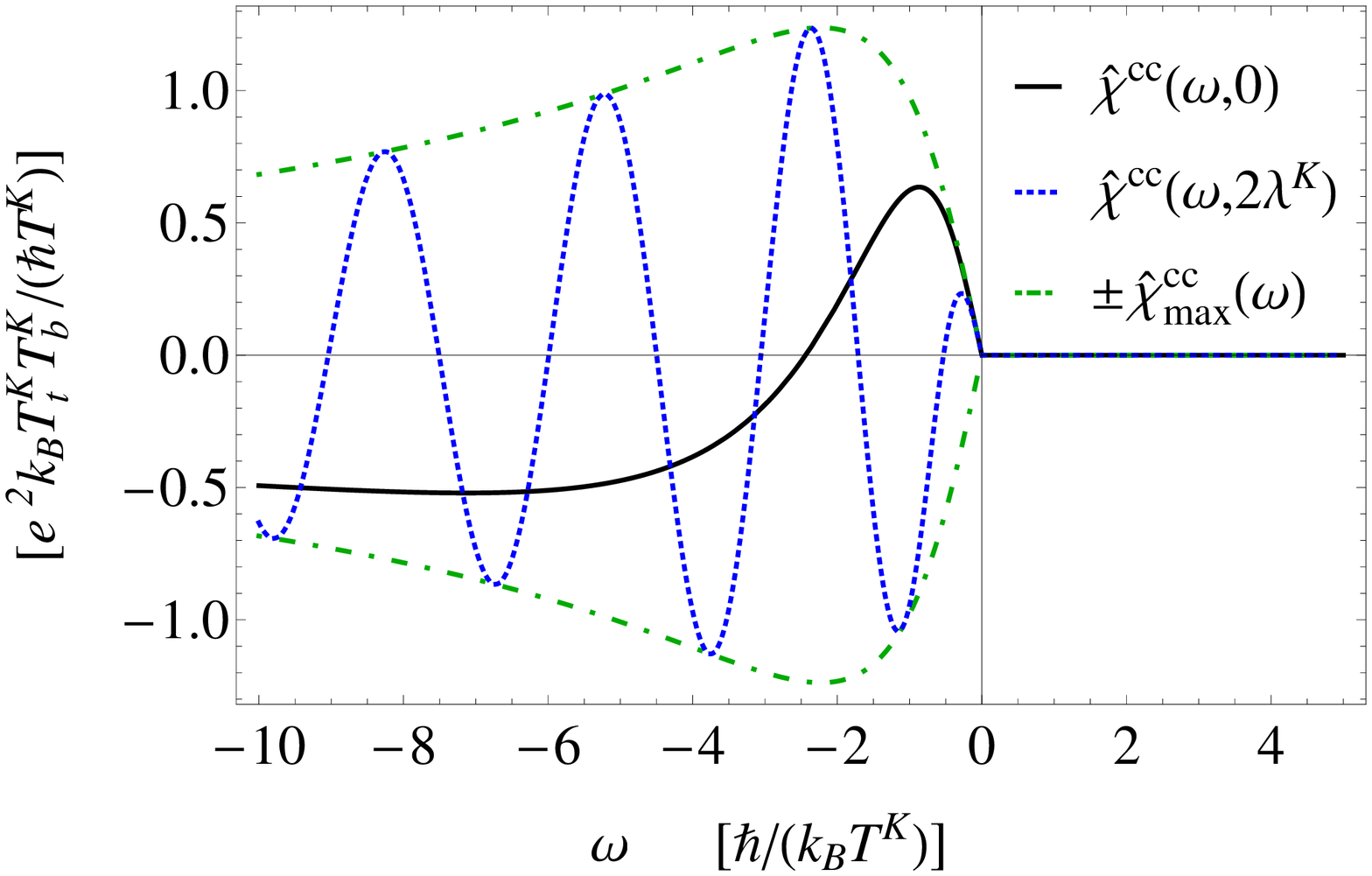}
}
\caption{\label{figFourierTransformsOfKCAndCCC}(Color online) Fourier transforms of the total Kondo cloud, top panel, and the current cross correlations, bottom panel, at zero temperature. 
The current cross correlations are depicted for the two representative values of $|x|+|y|$ equals zero and $|x|+|y|$ equals twice the Kondo length $\lambda^K$. Furthermore, we involve the envelope function $\hat{\chi}^{cc}_\text{max}$.
Despite the analogy of the Kondo cloud and the current cross correlations in the time domain (for certain time frames), they show no apparent similarity in the frequency domain.}
\end{figure}

To give a descriptive example of a manifestation of this relation, we derive the analytical formulas in the limit $j^z_\text{t/b} \to 0$ for both the Kondo cloud and the current cross correlations, following the lines of Ref. \onlinecite{PosskeLiuBudichTrauzettel2012ExactResultsForTheKondoCouldOfTwoHelicalLiquids}, and compare them.
For convenience, we introduce
\begin{align}
{\zeta}(x) = \frac{1}{\pi} e^{-\frac{\pi}{\hbar v \beta}x}\Phi
\left(
  e^{-\frac{2 \pi}{\hbar v \beta} x},1,\frac{1}{2} +\frac{\beta k_B T^K}{2 \pi} 
\right)	
\end{align}
that depends implicitly on the inverse temperature $\beta = 1/(k_B T)$
and the Kondo temperature \cite{SchillerHershfieldKondo}
\begin{align}
T^K = T^K_\text{t} + T^K_\text{b}.
\end{align}
Furthermore, $\Phi(z,s,a)$ is the Hurwitz-Lerch transcendent \cite{WolframHurwitzLerch}.
For the Kondo cloud, we obtain
\begin{align}
\chi^z_a(x) =  - \frac{\hbar k_B T^K_a}{2 v}  \zeta^2(|x|),
\label{eqConcreteExampleKondoCloud}
\end{align}
and the space- and time-resolved current cross correlations are
\begin{align}
\chi^{\text{cc}}(x,y,t) =&  
- \frac{e^2 k_B T^K_{\text{t}} k_B T^K_{\text{b}}}{\hbar^2}
\left\{
  \zeta^2\left(|x|+|y|-v t\right)^*
  \right.
  \nonumber \\
  &\left.
  + 
    \zeta^2\left(v t + |x| + |y|\right)
\right\}.
\label{eqConcreteExampleCCCorrelations}
\end{align}
Hence, the analytically derived formulas satisfy the proportionality relation of \refEq{eqFinalProportionality} and the proportionality factor is given by the corresponding Kondo temperature.
In experiments, it is common to measure frequency-resolved current cross correlations. 
We want to point out that the similarity in the time domain does
not simply transfer to the frequency domain. The reason is that  the full range of $t$ is taken into account in a Fourier transform and the proportionality in the time domain is limited to certain time frames.
To show this explicitly, we  present the Fourier transforms $\hat{f}(k) = \Pint dx \ e^{\I k x} f(x)$ of the Kondo cloud and the current cross correlations of Eqs.(\ref{eqConcreteExampleKondoCloud}) and (\ref{eqConcreteExampleCCCorrelations}) at zero temperature. Here, we take the principal-value Fourier transform since the Kondo cloud diverges at the site of the magnetic moment. The result is
\begin{align}
\hat{\chi}_a^z(k) =
- \frac{2 \hbar^2}{\pi^2} \frac{ T^K_a}{ T^K} 
\Re
\left\{
  \frac{ 
    \Li_2\left( 1 + \frac{\I \hbar v k}{k_B T^K} \right) + \frac{\pi^2}{12}
}
{2 + \frac{\I \hbar v k}{k_B T^K}}
\right\},
\label{eqFTKC}
\end{align}
where $\Li_2$ is the dilogarithm
and
\begin{align}
\hat{\chi}^{\text{cc}}(\omega,|x|+|y|) =
{
\Theta(-\omega)
\frac{ 8 e^2 k_B}{\pi \hbar}  
\frac{T^K_\text{t} T^K_{\text{b}}}{T^K} 
}
\nonumber
\\
\times
\Re
\left\{
 \frac{ \I 
    e^{-\I \frac{\omega}{v} \left( |x|+|y| \right)} 
     \ln \left(1+ \frac{\I \hbar \omega}{k_B T^K} \right)
}{2+\frac{\I \hbar \omega}{k_B T^K}}
\right\}
\label{eqFTCCC}
\end{align}
with $\Theta$ being the Heaviside function. 
Instead of depicting the Fourier transform of the Kondo cloud at zero temperature of a single side, we concentrate on the more universal total Kondo cloud 
$\chi^z = \chi^z_\text{t} + \chi^z_\text{b}$. The functional form of its Fourier transform at zero temperature is shown in \refFig{figFTKC}. 
An interesting feature is revealed by $\hat{\chi}^z(0) = -\frac{\hbar^2}{4}$, i.e., the spatial integral over the total Kondo cloud  in the ground state equals the expected value for exact screening.
The Fourier transform of the current cross correlations at zero temperature in turn is illustrated in \refFig{figFTCCC}. It is convenient to introduce the envelope function 
\begin{align}
\hat{\chi}^{cc}_\text{max} =
\Theta(-\omega)
\frac{ 8 e^2 k_B}{\pi \hbar}  
\frac{T^K_\text{t} T^K_{\text{b}}}{T^K} 
\left|
 \frac{
     \ln \left(1+ \frac{\I \hbar \omega}{k_B T^K} \right)
}{2+\frac{\I \hbar \omega}{k_B T^K}}
\right|.
\end{align}
For generic values of $|x|+|y|$, $\chi^{cc}(\omega,|x|+|y|)$ oscillates between $\pm \hat{\chi}^{cc}_\text{max}$. Hence, for clarity, we choose two representative values for $|x|+|y|$ in \refFig{figFTCCC}. First, we set $|x|+|y|$ twice the Kondo length $\lambda^K$, which is a typical value for the setup at hand in the sense that a broad range of the Kondo cloud can be spatially resolved in the time domain.
Secondly, we look at $|x|+|y|=0$.
Evidently, there is no obvious similarity between the Kondo cloud and the current cross correlations in the frequency domain. 

\section{Dependence of central result on interaction strengths \label{sctnDiscussion}}
The experimental realization of the proposed setup is highly challenging. 
Nevertheless, we have chosen to treat the system at hand because it shows the clearest possible appearance of the Kondo cloud in the current cross correlations.
This feature can be affected by altering the interaction strengths $g_\text{t/b}$, which we are going to discuss now. 
Allowing for an unrestricted choice of $g_\text{t/b}$ would exceed the scope of this work, since, in general, there exists no simplifying refermionization. 
Covered by the method at hand, however, are the two lines $g_\text{t} + g_\text{b}=1$ and $g_\text{t} + g_\text{b}=2$.
In both cases, correlators of the form $\langle \Psi^\dagger \tau^- \Psi^\dagger \Psi \rangle$ appear additionally in the current cross correlations, where the fermionic fields $\Psi$ can be of different leads, space, and time.
Although these additional correlators are interesting objects themselves, they hinder the direct measurement of the Kondo cloud in principle.

As the second line $g_\text{t} + g_\text{b} = 2$ is based on a different effective Hamiltonian \cite{PosskeLiuBudichTrauzettel2012ExactResultsForTheKondoCouldOfTwoHelicalLiquids}, we limit ourselves here to describing the first line $g_\text{t} + g_\text{b} = 1$ in greater detail. 
If we leave $g_\text{t} = g_\text{b} = \frac{1}{2}$ with the constraint $g_\text{t}+g_\text{b}=1$, the Hamiltonian of \refEq{eqSimplifiedHamiltonian} slightly changes, so that all appearing fields $\Psi_\text{t}$ are replaced by $\Psi_4$ and all fields $\Psi_\text{b}$ are replaced by $\Psi_2$ 
with the relations
\begin{align}
\left(
\begin{array}{c}
  \rho_2
\\
  \rho_4
\end{array}
\right)
= 
\frac{1}{\sqrt{2}}
\left(
\begin{array}{cc}
{\scriptstyle
\sqrt{g_\text{t}}- s \sqrt{g_\text{b}}	}&	{\scriptstyle - s \sqrt{g_\text{t}} - \sqrt{g_\text{b}}}
\\
{\scriptstyle-\sqrt{g_\text{t}} -s \sqrt{g_\text{b}}}	&	{\scriptstyle -s \sqrt{g_\text{t}} + \sqrt{g_\text{b}}}	
\end{array}
\right)
\left(
\begin{array}{c}
  \rho_{\text{t}}
\\
  \rho_{\text{b}}
\end{array}
\right),
\end{align}
where $s \in \{-1,+1\}$.
The resulting transformed Hamiltonian is
\begin{align}
\mathcal{H}^\perp_{K} &= \hat{j}^\perp_\text{t} \Psi_\text{4}^\dagger(0)\tau^- + \hat{j}^\perp_\text{b} \Psi_\text{2}^\dagger(0)\tau^- + H.c.,
\nonumber
\\
\mathcal{H}^z_{K} &= \hat{j}_\text{t}^z \Psi^\dagger_\text{4}(0) \Psi_\text{4}(0) \tau^z + \hat{j}_\text{b}^z \Psi^\dagger_\text{2}(0)\Psi_\text{2}(0) \tau^z 
\label{eqSimplifiedHamiltonian2}
\end{align}
with $\hat{j}^\perp_a = j^\perp_a$ and $\hat{j}^z_a = \frac{j^z}{\sqrt{2 g_a}} - \sqrt{2} \pi v(\sqrt{g_a} + s\sqrt{g_{\lnot a}})$ where $\lnot\text{t} = \text{b}$ and $\lnot\text{b} = \text{t}$.
Note that the calculations done in first order around $j^z_a = 0$ in Section \ref{sctnResults} now hold in first order around $\hat{j}^z_a = 0$, which corresponds to $j^z_a = 2 \pi v(g_a + s\sqrt{g_a g_{\lnot a}})$. 
We want to put special emphasis on the fact that, for $s=+1$, the couplings $j^z_\text{t/b}$ do not need to be small compared to $j_\text{t/b}^\perp$, which could be seen as unphysical considering that the respective bare couplings should be of the same order. In fact, to assume a small ratio $j^z_\text{t/b} / j^\perp_\text{t/b}$ is not unphysical as the renormalization group (RG) analysis given in Ref. \onlinecite{PosskeLiuBudichTrauzettel2012SupplementalMaterial} shows. Since the terms coupled to $j^\perp_\text{t/b}$ are relevant while the terms coupled to $j^z_\text{t/b}$ are part of the free Hamiltonian in the RG calculations, $j^\perp_\text{t/b}$ initially grows following the RG flow. Hence, starting with small bare couplings of the same order, the RG flow generates a situation with a small ratio $j^z_\text{t/b} / j^\perp_\text{t/b}$.
For the current cross correlations, we obtain the additional contribution
 \begin{align}
 \chi^{cc}_{AS} &=
 \sum_{\sigma,\sigma^\prime \in \{\pm\}}
	s \frac{e^2}{\hbar}
  \frac{g_\text{b} - g_\text{t}}
  {4\sqrt{g_\text{t}g_\text{b}}}
	\nonumber
	\\
  &\times \left\langle
	\delta \rho_4(\sigma x) \delta \rho_4(\sigma^\prime y, t) 
   \right.
		\left.
    -
    \delta \rho_2(\sigma x) \delta \rho_2(\sigma^\prime y, t) 
		\right\rangle
\label{eqAsymmetricCurrentCurrentContribution}.
\end{align} 
Terms of the form  $\langle \Psi^\dagger \tau^- \Psi^\dagger \Psi \rangle$ appear here by applying \refEq{eqBackpropagationOfDensity} to the density operators with positive spatial argument.

The quantity $\chi^{cc}_{AS}$ is anti-symmetric in both pairs of couplings $(g_\text{t},g_\text{b})$ and $(\hat{j}_\text{t},\hat{j}_\text{b})$.
The latter is seen by considering the invariance exhibited by the Hamiltonian in \refEq{eqSimplifiedHamiltonian2} under simultaneous exchange of the fields $\Psi_4 \leftrightarrow \Psi_2$ and exchange of the couplings $\hat{j}_\text{t} \leftrightarrow \hat{j}_\text{b}$.
In this regard, $\chi^{cc}_{AS}$ vanishes
for an equal coupling to the magnetic moment $\hat{j}_\text{t} = \hat{j}_\text{b}$,
but, in general, we encounter a perturbation of the proportionality between the Kondo cloud and the current cross correlations.%

However, $\chi^{cc}_{AS}$ can be eliminated under the assumption that results of the crossed current correlations for several values of the couplings $\hat{j}$ are available
A concrete example for an elimination is to add up the current cross correlations of two systems, where the second system differs from the first one only by exchanged tunnel couplings $\hat{j}_\text{t} \leftrightarrow \hat{j}_\text{b}$. By the above-mentioned invariance of the Hamiltonian, this exchange is equivalent to the exchange of $\rho_4 \leftrightarrow \rho_2$.
The proportionality relation of \refEq{eqAnalyticallyProportionality} is then only slightly altered. Instead of the Kondo cloud of one lead, a linear combination of the Kondo clouds of both leads occurs.
For instance, for $(|x| + |y|)/v > t > \tau^K$ and first order in ${\hat{j}}^z$ we obtain
\begin{align}
\chi^{cc}(x,y,t) + \chi^{cc}_{\hat{j}_\text{t} \leftrightarrow \hat{j}_\text{b}}(x,y,t)
=
c_\text{t} \chi^z_\text{t} + c_\text{b} \chi^z_\text{b} 	
\end{align}
with
\begin{align}
c_\text{t} =  -\frac{2 e^2 v k_B}{\hbar^3} \left(s \sqrt{g_\text{t} g_\text{b}} T^K + g_\text{t} T^K_M \right)
\\
c_\text{b} = -\frac{2 e^2 v k_B}{\hbar^3} \left(s \sqrt{g_\text{t} g_\text{b}} T^K - g_\text{b} T^K_M \right),
\end{align}
where we introduced the auxiliary temperature $T^K_M = T^K_\text{t} - T^K_\text{b}$.
\section{Summary \label{sctnSummary}}
We have argued for a generic connection between the Kondo cloud and the space- and time-resolved current cross correlations in a Kondo setup with helical liquid leads. The relation relies on the equality of current and spin-density in the helical liquid. 
For the case of two helical liquid leads with Luttinger parameters $g_\text{t}=g_\text{b}=1/2$, the relation is a direct proportionality up to first order in the $z$-spin coupling if the measurement is taking place sufficiently far away from the magnetic moment.
This requirement is a merit rather than a restriction, because a distant manipulation of the system is incapable of directly affecting the magnetic moment.
With respect to experimental realizations, we show that it is required to be able to measure the current cross correlations in the time domain. Furthermore, we discuss disturbances if the preferred point of interactions, i.e., $g_\text{t} = g_\text{b} = \frac{1}{2}$, is left and give ideas of how to restore the proportionality.
As a consequence of our results, probing the current cross correlations describes a tool to directly detect the Kondo cloud and even resolve it spatially. The associated measurements are conceptually more promising than a direct measurement of the correlation between the magnetic moment and the spin density in the leads because the magnetic moment is in no manner directly perturbed.   
%

We acknowledge discussions with Jan Budich, Fabrizio Dolcini, and Michele Filippone
and financial support by the DFG (SPP1666 and the DFG-JST research unit ''Topotronics``) as well as the Helmholtz Foundation (VITI).

\setcounter{equation}{0}
\appendix*
\section*{Appendix \label{appendix}}
In this appendix, we calculate the time evolution of the density operators of the Hamiltonian in \refEq{eqSimplifiedHamiltonian}:
\begin{align}
\mathcal{H}^0 &= 
\sum_{j \in \{\text{t},\hat{\text{t}},\text{b},\hat{\text{b}}\}} 
\int dx \ 
\Psi^\dagger(x)_{j}(- \I \partial_x) \Psi(x)_{j} 	
,
\nonumber
\\
\mathcal{H}^\perp_{K} &= j^\perp_\text{t} \Psi_\text{t}^\dagger(0)\tau^- + j^\perp_\text{b} \Psi_\text{b}^\dagger(0)\tau^- + H.c.,
\nonumber
\\
\mathcal{H}^z_{K} &= j_\text{t}^z \Psi^\dagger_\text{t}(0) \Psi_\text{t}(0) \tau^z + j_\text{b}^z \Psi^\dagger_\text{b}(0)\Psi_\text{b}(0) \tau^z,
\end{align}
setting $\hbar = v = 1$.
In particular, we focus on evolving $\rho(x)$ with $x>0$ backwards in time to shortly before it has interacted with the impurity. For this reason, we consider the operator
\begin{align}
\Psi(x,-x-\epsilon)=e^{-\I \mathcal{H} (x+\epsilon)}\Psi(x,t=0) e^{\I \mathcal{H} (x+\epsilon)},
\end{align}
with $\epsilon \in \mathbb{R}^+$, where we abandon the index of the lead as the calculation is valid for either lead.
To compute the time evolution, we look at the time derivative of $\Psi$. For spatial arguments that are not equal to zero, the time evolution just becomes the linear propagation because the interaction with the magnetic moment is localized at $x = 0$. 
A technical problem in this procedure appears directly at $x=0$, where we would have to evaluate the commutator $\left[\Psi(0,0), H \right]$ which is not well defined. The reason for this ''divergence`` of the derivative of the fermionic field is, physically speaking, the vanishing size of the impurity.
To avoid this type of divergence, we introduce the smeared delta function $\delta_d^y(x)$ that is centered at $y \in \mathbb{R}$. The exact shape of this function is unimportant. Of importance are the following properties that are going to be used later:
\begin{align}
\lim_{d \to 0}\int_{I\subset\mathbb{R}} dx \ \delta^y_d P(x) =
\begin{cases}
P(y) \ \text{if $y \in \mathring{I} \subset \mathbb{R}$},
\\
\frac{1}{2} P(y) \ \text{if $y \in \partial I \subset \mathbb{R}$},
\\
0 \ \text{else},
\end{cases}
\label{eqSmearedDeltaProperty1}
\end{align}
for $P$ a sum of products of the fields $\tau^+$, $\tau^{-}$, $\Psi$, and $\Psi^\dagger$ in the Heisenberg picture, where $x$ can appear in the time argument as well; $\mathring{I}$ denotes the interior of $I$ and $\partial I$ denotes the boundary of $I$.
Furthermore, the following identity has to hold for multidimensional integrals
\begin{align}
&\lim_{d \to 0} \int dx_1 \dots dx_n \ \delta^{y_1}_{d}(x_1) \dots \delta^{y_n}_{d}(x_n) P(\vec{x})
=
\nonumber
\\
&\lim_{d_1\to 0} \dots \lim_{d_n \to 0} \int dx_1 \dots dx_n \ \delta^{y_1}_{d_1}(x_1) \dots \delta^{y_n}_{d_n}(x_n) P(\vec{x}).
\label{eqSmearedDeltaProperty2}
\end{align}
A concrete example of the smeared delta function is the Lorentzian
\begin{align}
\delta^y_d(x) = \frac{d}{\pi(d^2+(x-y)^2)},
\end{align}
so that the length scale $d$ can be interpreted as the size of the impurity.
The interaction Hamiltonians with a broadened impurity become
\begin{align}
\mathcal{H}^{\perp}_{K} &=
\sum_{a\in\{\text{t},\text{b}\}} j^\perp_a \int dx \delta_d^0(x)\Psi^\dagger_a(x)  \tau^-      \
+ H.c.,
\nonumber
\\
\mathcal{H}^z_K &= \sum_{a\in\{\text{t},\text{b}\}} j^z_a \int dx  \delta^0_d(x) \Psi^\dagger_a(x) \Psi_a(x) \tau^z.
\end{align}
The time derivative of $\Psi(x)$ is then given by
\begin{align}
\partial_t \Psi(x,t) |_{t=0} =\ & 
\I \left[ \Psi(x,0) , \mathcal{H} \right] 
\nonumber 
\\
=\ & \partial_x \Psi(x) + \I j^\perp \delta^x_d(0) \Psi(x) \tau^+ 
\nonumber
\\
&+ \I j^z \delta^x_d(0) \Psi(x) \tau^z.
\end{align}
We now solve the time evolution for the operator $\Psi$ exactly by cutting the time into infinitesimally small slices, solve the evolution of a time slice exactly, and iterate.  This technique shows similarities to an approach taken in the derivation of the path integral for quantum mechanics\cite{altland2006condensed}.
We define $\epsilon\gg d$.
Then, we obtain
\begin{align}
e^{-\I \mathcal{H}2 \epsilon} \Psi(\epsilon,0) e^{\I \mathcal{H} 2 \epsilon}
=
\lim_{N \to \infty} \left(1 - \I \mathcal{H} \eta \right)^N \Psi(\epsilon,0) \left(1+ \I \mathcal{H} \eta \right)^N
\end{align}
with $\eta = \frac{2 \epsilon}{N}$, where the convergence of the right hand side towards the left hand side is proven by comparing all orders in $\epsilon$.
Denoting $ \left(1 - \mathcal{H} \eta \right)^k\Psi(\epsilon,0) \left(1 + \I \mathcal{H} \eta \right)^k =: \Psi^k(\epsilon,0)$, we look at the first time slice
\begin{align}
\Psi^1(\epsilon,0) =\ & \Psi(\epsilon,0) - \I \eta \left[\mathcal{H}, \Psi(\epsilon,0)\right] + \mathcal{O}(\eta^2),	
\nonumber
\\
=\ & 
\Psi(\epsilon,0) - \eta \partial_x \Psi_a(\epsilon) - \I j_a^\perp \delta^\epsilon_d(0) \tau^-(0) 
\nonumber
\\
&- \I j_a^z \delta^\epsilon_d(0) \tau^z(0) \Psi(0,0)+ \mathcal{O}(\eta^2) 
\nonumber
\\
=\ &
\Psi(\epsilon-\eta,0) - \I j_a^\perp \delta^\epsilon_d(0) \tau^-(0)
\nonumber
\\
& - \I j_a^z \delta^\epsilon_d(0) \Psi_a(0) \tau^z(0) + \mathcal{O}(\eta^2), 
\label{eqTimeEvolutionPsi1}
\end{align}
where we used $\left[\Psi,\tau^z\right]=0$.
Since we take $\eta \to 0$  at the end of our calculations, we henceforth neglect terms
$\mathcal{O}\left(\eta^2\right)$.
This is justified by the identity
\begin{align}
\lim_{N\to\infty} \left(1+\frac{x}{N}+\mathcal{O}(\eta^2) \right)^N = e^x = \lim_{N\to\infty} \left(1+\frac{x}{N} \right)^N.
\end{align}
\noindent
The general formula for an arbitrary number of time steps is given by
\begin{align}
\Psi^k =\ &
\Psi(\epsilon - \eta k)
\nonumber
\\
&
+ \sum_{n=1}^{\infty} {(- \I j^z \eta)}^n \sum_{(l_i)_{i = 1}^n>0}^{\sum_{i=1}^n l_i \leq k-n} \delta^\epsilon_d(\eta l_1)
\left[
\prod_{j=2}^n \delta^0_d(\eta l_j)
\right]
\nonumber
\\
&
\times
\left[
\prod_{j=1}^n \tau^z\left(\eta\left({n-k - j + \sum_{i=1}^{n-j} l_i}\right)\right)
\right] 
\nonumber
\\
&
\times
\left(
  \frac{j^\perp}{j^z} \tau^-(\eta \kappa) + \tau^z(\eta \kappa) \Psi(\eta \kappa,0)
\right)\Biggm|_{\kappa=n-k + \sum_{i=1}^n l_i}.
\end{align}
This can be proven by complete induction over $k$.
Here, $(l_i)_{i = 1}^n>0 $ denotes all $n$-tuples of integers, each of which is larger than $0$, and the product assumes an ordering of its factors from the left to the right with increasing index, i.e., $\prod_{j=1}^n a_j = a_1 \times \dots \times a_n$. 
To obtain the continuum limit in time and hence the time evolution, we let $N$ go to infinity, which is equivalent to $\eta \to 0$. In doing so, we replace $\eta N \to 2 \epsilon$ and set all products of the form $\eta \times c$  to zero, with $c$ being a fixed number.
Furthermore, sums of the form $\sum_{l=1}^N \eta f(\eta l)$ become integrals for non diverging functions and operators $f$.
This procedure yields
\begin{widetext}
\begin{align}
\Psi(\epsilon, -2 \epsilon) =\ & \Psi(-\epsilon,0) + 
%
%
\sum_{n=1}^\infty
(- \I j^z_a)^n
\int_0^{2 \epsilon} dl_1
\int_0^{2 \epsilon-l_1} dl_2
\dots
\int_0^{2 \epsilon - (\sum_{i=1}^{n} l_i)} dl_n
\delta^\epsilon_d(l_1)
\left[
\prod_{j=2}^n
\delta^0_d(l_j)
\right]
\nonumber
\\
&\times
  \left[
  \prod_{j=1}^{n}
  \tau^z(-2\epsilon + \sum_{i=1}^{n-j} l_i )
  \right]
\left(
  \frac{j^\perp}{j^z} \tau^-(-2\epsilon+\sum_{i=1}^n l_i) + 
  \tau^z(-2\epsilon+\sum_{i=1}^n l_i)
  \Psi(-2\epsilon+\sum_{i=1}^n l_i,0) 
\right)
.
\end{align}
\end{widetext}
For a particular shape of the impurity (encoded in $\delta_d$), the integrals in general do not simplify further. But by taking the limit $d \to 0$, belonging to the case that the size of the impurity is considerably smaller than the resolution of the possible measurements, we can exploit the properties of the representation of the smeared delta function postulated in \refEq{eqSmearedDeltaProperty1} and \refEq{eqSmearedDeltaProperty2} to obtain
\begin{align}
\Psi(\epsilon,-2 \epsilon) 
=\ &  \Psi(-\epsilon,0) \frac{1-(\frac{j^z}{4})^2}{1+(\frac{j^z}{4})^2} - \tau^-(-\epsilon) \frac{\I j^\perp}{1+ \I \frac{j^z}{4}} 
\nonumber
\\
&
- \tau^z(-\epsilon) \Psi(-\epsilon,0)  \frac{\I j^z}{1+(\frac{j^z}{4})^2}.
\label{eqPsiFieldTimeBackShiftFinal}
\end{align}
In this derivation, the property $\left(\tau^z \right)^2 = \frac{1}{4}$, and the identity $\sum_{n=0}^\infty (\I \frac{j^z}{4})^n = \frac{1}{1- \I \frac{j^z}{4}}$ for $|j^z|<4$ have been exploited.
Starting from \refEq{eqPsiFieldTimeBackShiftFinal}, the evolution in time for the density operator becomes
\begin{align}
\rho(\epsilon,-2 \epsilon) =\ & \Psi^\dagger\Psi(\epsilon,-2\epsilon) 
\nonumber
\\
=\ & \rho(-\epsilon,0)  + \frac{\left( j^\perp \right)^2}{1+(\frac{j^z}{4})^2} \left( \tau^z(-\epsilon)+1/2 \right)
\nonumber
\\
&
+ \left( \I j^\perp \frac{1+\I \frac{j^z}{4}}{(1-\I \frac{j^z}{4})^2} \tau^+(-\epsilon) \Psi(-\epsilon,0) + H.c.\right) .
\end{align}
Notably, the term  containing the operators $\Psi^\dagger(-\epsilon,0) \tau^z(-\epsilon) \Psi(-\epsilon,0)$ vanishes because its pre\-fac\-tor is zero.

\clearpage
\bibliography{BibMeasureTheKondoCloud}

\end{document}